\documentclass[aps,prl,twocolumn,groupedaddress,showpacs]{revtex4-1}
\usepackage[pdftex]{graphicx}
\usepackage{epstopdf}

\begin{document}


\title{Local Compressibility Measurements of Correlated States in Suspended Bilayer Graphene}


\author{Martin, J.}
\author{Feldman, B. E.}
\author{Weitz, R. T}
\author{Allen, M. T.}
\author{Yacoby, A.}
\affiliation{Department of Physics, Harvard University, Cambridge, MA 02138, USA}


\date{\today}

\begin{abstract}
Bilayer graphene has attracted considerable interest due to the important role played by many-body effects, particularly at low energies.  Here we report local compressibility measurements of a suspended graphene bilayer.  We find that the energy gaps at filling factors $\nu = \pm4$ do not vanish at low fields, but instead merge into an incompressible region near the charge neutrality point at zero electric and magnetic field.  These results indicate the existence of a zero-field ordered state and are consistent with the formation of either an anomalous quantum Hall state or a nematic phase with broken rotational symmetry.  At higher fields, we measure the intrinsic energy gaps of broken-symmetry states at $\nu = 0, \pm 1$ and $\pm 2$, and find that they scale linearly with magnetic field, yet another manifestation of the strong Coulomb interactions in bilayer graphene.
\end{abstract}

\pacs{73.22.Pr, 73.43.-f}
\keywords{graphene, quantum Hall effect}

\maketitle


The low-energy dispersion of bilayer graphene can be described to first order by parabolic valence and conduction bands that meet at the Fermi energy \cite{castro2009}.  The charge excitations of this band structure are massive chiral fermions whose quantum Hall signature is different from that of both monolayer graphene \cite{novo2005,zhang2005} and conventional two-dimensional electron gases.  In the absence of interactions, the Landau level (LL) energy spectrum of bilayer graphene is given by $E_N = \pm \hbar \omega_c \sqrt{N(N-1)}$, where $N$ is the orbital index,  $\hbar = h/2 \pi$ ($h$ is Planck's constant), $\omega_c = eB/m^\ast$ is the cyclotron frequency, $e$ is the electronic charge, $B$ is the magnetic field, and $m^\ast$ is the effective mass \cite{novo2006}.  Plateaus occur in Hall conductivity at $\sigma_{xy} = 4Me^2/h$, where the factor of four is due to spin and valley degeneracy and $M$  is a nonzero integer, reflecting the eightfold-degenerate LL at zero energy, which is comprised of $N = 0$ and $1$ orbital states.
\\
\indent
When magnetic field is large enough or disorder is sufficiently low, interaction effects such as quantum Hall ferromagnetism (QHF) \cite{barlas2008} or magnetic catalysis \cite{ezawa2007} are predicted to open energy gaps and give rise to additional plateaus in the quantum Hall spectrum at intermediate filling factors.  Recent transport measurements have indeed revealed signatures of many-body effects in bilayer graphene \cite{feldman2009,zhao2010,bao2010,dean2010} and the dependence of resistance on temperature and magnetic field was used to determine the magnitude of the interaction-driven energy gaps \cite{feldman2009,zhao2010}.  More recently, it has been theoretically predicted that spontaneously broken symmetries will occur in bilayer graphene at zero magnetic field.  The nature of the zero-field interacting phase is still under intense theoretical debate, with suggestions of spontaneous transfer of charge between layers \cite{min2008,zhang2010} ferroelectric domains \cite{nandprl2010}, nematic ordering \cite{vafek2010,lemonik2010}, or the formation of an anomalous Hall insulator \cite{haldane88,raghu2008,min2008,nand2010,zhang2010,rahul2010}.
\\
\indent
One way to distinguish between the various interacting states is to determine whether an energy gap is present at $B = 0$.  Transport measurements can provide an indication of gap size, but it is known that disorder can decrease the apparent transport gap relative to the true intrinsic gap \cite{oostinga2008,zhang2009}.  Therefore, it is desirable to directly probe electronic properties in a thermodynamic measurement.  Here, we report the use of a scanning single-electron transistor (SET) \cite{yoo97,yacoby99,martin2008,martin2009} to measure the local compressibility of a suspended bilayer graphene flake.  Investigations into the compressibility of bilayer graphene were recently reported on unsuspended samples \cite{henriksen2010,young2010} with particular attention paid to the gap induced by an electric field, but disorder in these systems was too large to observe the broken-symmetry states discussed above.  Our measurement combines the high sensitivity afforded by an SET with the low disorder of suspended devices, allowing us to study electronic states that arise from Coulomb interactions and revealing the existence of an ordered state at zero field.  A schematic illustration of the measurement system is shown in Fig. 1(a); for a full description of the measurement technique, see refs. \cite{yoo97,yacoby99,supplement}.  The SET is capable of measuring changes in local electrostatic potential $\Phi$ with $\mu V$ sensitivity.  As carrier density $n$ is varied, changes in $\Phi$ directly reflect the changes in local chemical potential $\mu$ of the bilayer flake, so the scanning SET tip can be used as a local probe of inverse compressibility (incompressibility) $d\mu/dn$.
\begin{figure}
\includegraphics[width=75mm]{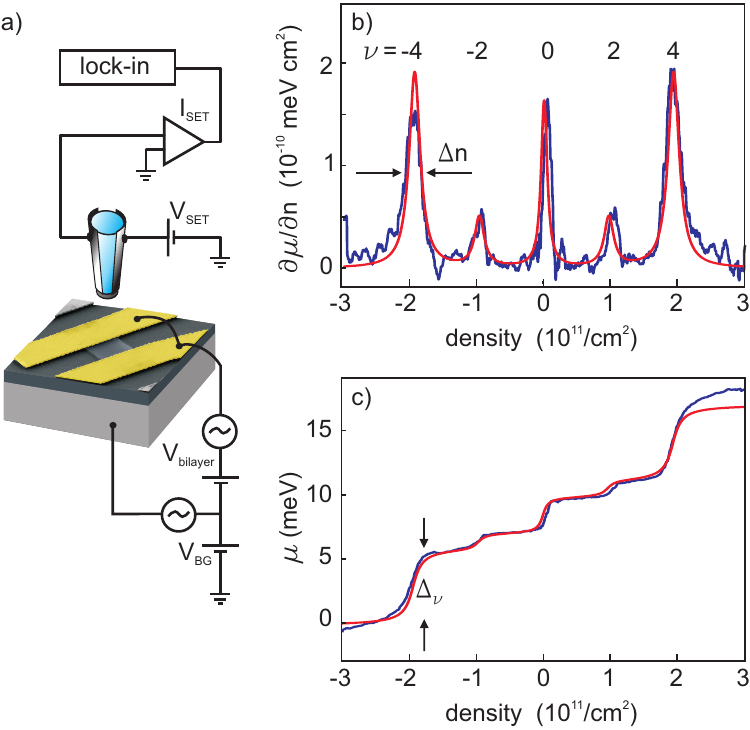}%
\caption{(Color online) (a) Schematic illustration of the measurement setup.  (b) Inverse compressibility as a function of carrier density at $B = 2$ T. Incompressible peaks at $\nu = 0, \pm 2,$ and $\pm4$ occur due to the decreased density of states between Landau levels.  (c) Chemical potential as a function of carrier density at $B = 2$ T, obtained by integrating the data in b. Steps in chemical potential occur at $\nu = 0, \pm 2,$ and $\pm 4$ as electrons begin to occupy the next Landau level.  In panels b and c, experimental data are shown in blue and fits, based on Lorentzians at each filling factor, are shown in red.\label{fig1}}
\end{figure}

In order to establish our measurement technique, we first describe the behavior of our sample in the high-field regime.  Figure 1(b) shows a typical measurement of inverse compressibility taken at $B = 2$ T.  Peaks in $d\mu/dn$, caused by the low density of states between neighboring LLs, are apparent at $\nu = 0, \pm2$ and $\pm4$.  The observation of incompressible regions at $\nu = 0 $ and $ \pm2$ is indicative of broken symmetries in the zero-energy LL, consistent with recent transport measurements \cite{feldman2009,bao2010,zhao2010,dean2010}.  The widths of the incompressible regions provide a measure of disorder \cite{martin2009}, and the full width at half maximum at $\nu = 4$ is on the order of $10^{10}$ cm$^{-2}$.  This is more than ten times smaller than in unsuspended devices \cite{morozov2008}, and is consistent with estimates from transport measurements of similar suspended bilayers \cite{feldman2009}.
\\
\indent
Chemical potential as a function of carrier density, shown in Fig. 1(c) for $B = 2$ T, is obtained either by direct measurement or by integrating curves similar to that shown in Fig. 1(b).  The steps in $\mu(n)$ at each filling factor provide a measure of the energy gaps $\Delta_{\nu}$ between neighboring LLs.  We have measured gap size at $\nu = 0, \pm 1, \pm 2, \pm 4$ and $\pm 8$ as a function of magnetic field, and the resulting data are shown in Fig. 2(a).  Consistent with the expected behavior of $E_N, \Delta_4$ and $\Delta_8$ scale linearly with magnetic field, with magnitudes of $3.9$ meV/T and $2.8$ meV/T, respectively.  This linear dependence confirms that the range of carrier densities probed in our experiment lies within the energy regime where the bilayer graphene band structure is well-approximated by parabolic bands.  It should be contrasted with cyclotron resonance and compressibility studies performed at higher densities where the hyperbolic nature of the dispersion was apparent from the sublinear dependence of gap size on magnetic field \cite{henriksen2010,henriksen2008}.
\begin{figure}
\includegraphics[width=60mm]{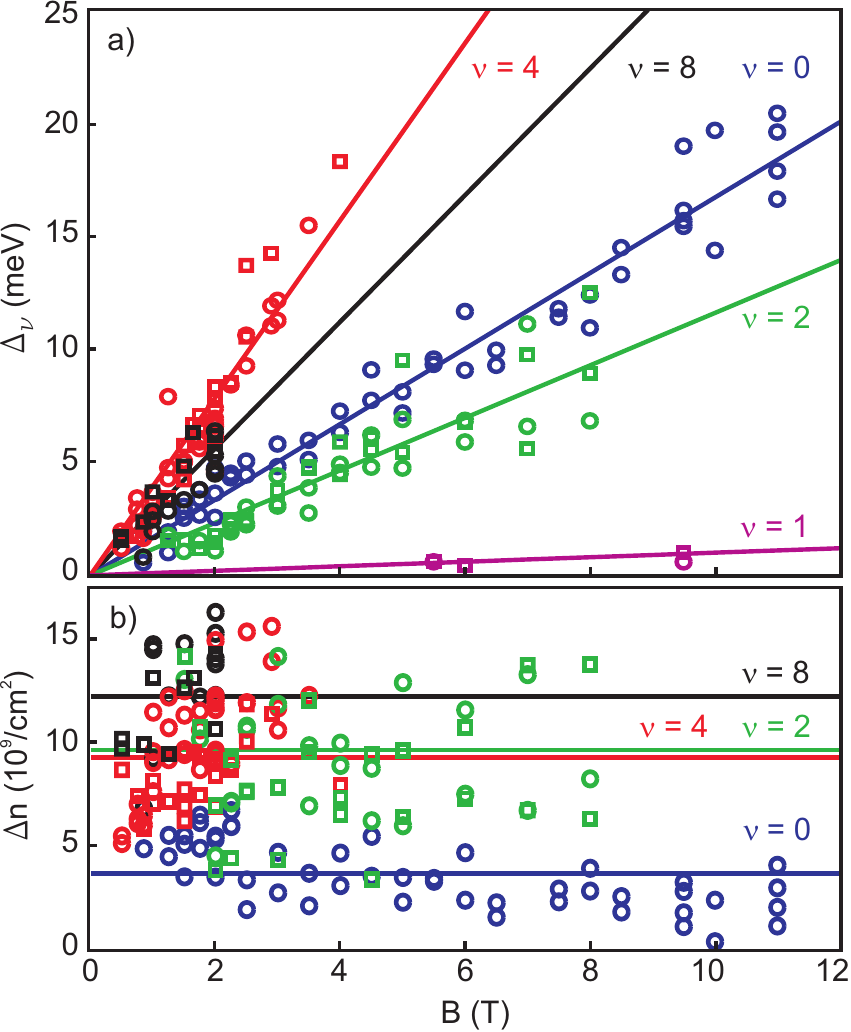}%
\caption{(Color online) (a) Gap size, extracted from a Lorentzian fit, as a function of magnetic field for $\nu = 0$ (blue), $\pm 1$ (magenta), $\pm 2$ (green), $\pm 4$ (red), and $\pm 8$ (black). Squares correspond to electrons, and circles to holes.  Gap size at all filling factors is well described by linear scaling with magnetic field, with fits given by the solid lines.  (b) Incompressible peak width as a function of magnetic field for $\nu = 0$, $\pm 2$, $\pm 4$, and $\pm 8$ (same colors as in a).  The $\nu = 0$ peak is significantly narrower than the others.  Peak width does not strongly depend on magnetic field.}
\end{figure}

It is apparent from Fig. 2(a) that the energy gaps $\Delta_0$ and $\Delta_2$ corresponding to broken-symmetry states at $\nu = 0$ and $\pm 2$ also increase proportionally with $B$, and have gap sizes of $1.7$ meV/T and $1.2$ meV/T, respectively.  For $\nu = 0$, the data show good agreement with the activation energy $E_a = \Delta_0/2 = 0.3-0.9$ meV/T measured in transport experiments \cite{feldman2009}, as well as recent theoretical predictions \cite{nandprl2010,gorbar2009,gorbar2010}.  For $\nu = 2$, activation experiments on unsuspended flakes \cite{zhao2010} yielded significantly smaller gap sizes and suggested a $B^{1/2}$ dependence for the gap, which seems to conflict with our measurement.  However, the error bars from the transport measurement are large enough that a linear fit to the data which passes through the origin is not inconceivable.  Linear scaling with $B$ for interaction-driven LLs is reasonable if one considers screening from higher orbital LLs, whose energy separation is much smaller than the Coulomb energy for all experimentally relevant fields \cite{nandprl2010}.  The linear scaling of $\Delta_0$ and $\Delta_2$ can therefore be understood as a result of the very strong Coulomb interactions in bilayer graphene.
Finally, our sample exhibits energy gaps of less than $1$ meV at $\nu = \pm 1$ in high magnetic fields.  Linear scaling with a slope of approximately $0.1$ meV/T provides a reasonable fit to these data, but the gap sizes are too small to conclusively rule out $B^{1/2}$ dependence.
\\
\indent
We can use the measured gap sizes to determine the effective mass of bilayer graphene.  Summing the measured energy gaps for $|\nu | \le 4$ at each magnetic field and comparing with our expression for $E_N$, we obtain an estimate that $m^{\ast} = (0.027 \pm .002)m_e$, in good agreement with Shubnikov de Haas measurements \cite{castrev2007}.  From a similar analysis using the gap sizes for $|\nu | \le 8$, we extract an effective mass of $(0.032 \pm .002)m_e$, which is somewhat larger than the above value.  The origin of this discrepancy is unclear, but it may be an artifact of our analysis procedure \cite{supplement}.  It is also important to note that for samples with a single gate, density and electric field cannot be controlled independently, so all gaps at nonzero filling factors are measured in an electric field.  Changes in the expected effective mass could result from deviations from the parabolic band structure or from an electric field applied perpendicular to the flake \cite{castrev2007,mccann2006}.
\\
\indent
Figure 2(b) shows the widths of the incompressible regions at each filling factor as a function of magnetic field.  The peak at $\nu = 0$ is significantly narrower than the others [see also Fig. 1(b)], which may indicate that higher filling factors are subject to additional sources of disorder that do not affect the $\nu = 0$ state.  One possible explanation for this finding is the existence of variations in the effective magnetic field, which can be caused by ripples or strain \cite{giesbers2007,kats2008,kail2009}.
\begin{figure}
\includegraphics[width=85mm]{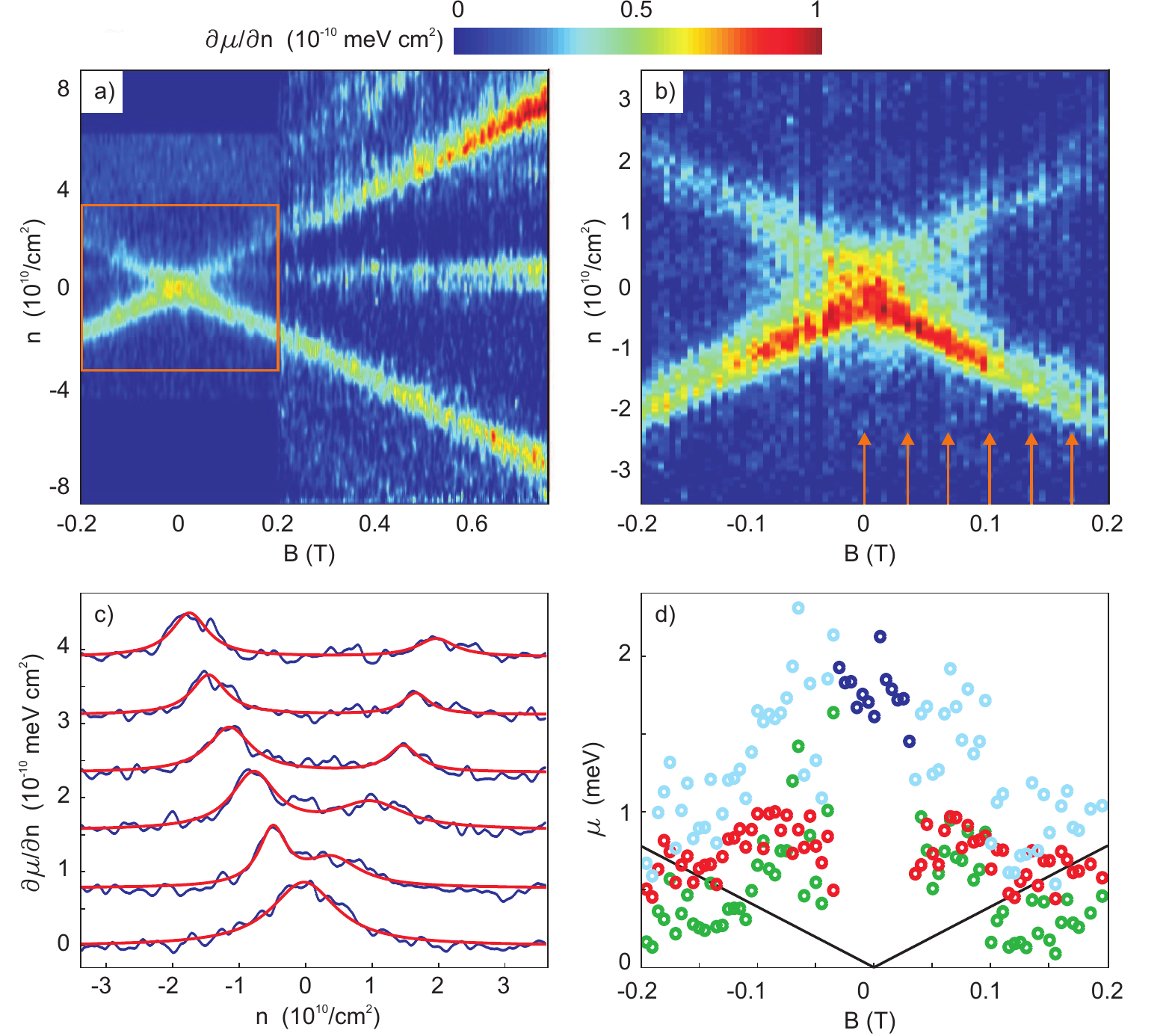}%
\caption{(Color online) (a) Inverse compressibility as a function of carrier density and magnetic field.  Incompressible peaks occur at $\nu = 0$ and $\pm 4$.  Below about $0.2$ T, peak height along $\nu = \pm 4$ increases with decreasing field, culminating in an incompressible peak at zero field.  (b), Zoom-in on the low-field behavior of the sample, taken at a different location on the flake from that in a.  Distinct peaks along $\nu = \pm 4$ persist all the way to zero field. (c) Line cuts of incompressibility along the red arrows shown in panel b for magnetic fields between $B = 0$ and $0.175$ T in steps of $0.035$ T.  Curves are offset for clarity.  Data are shown in blue, and the red curves are two-Lorentzian fits, except for the zero-field fit, which is composed of only one Lorentzian. (d) Gap size at $|\nu | = 4$ for electrons (green), holes (red), and their sum (cyan).  Data for $|B| > 0.03$ T is based on a two-Lorentzian fit.  The blue circles at low field describe the jump in chemical potential across the charge neutrality point, as modeled by a single Lorentzian fit.  Solid black lines correspond to $\Delta_4 = 3.9(B$[T]) meV, as derived from the data at high magnetic fields.}
\end{figure}

We now discuss the behavior of our sample at small magnetic fields.  Figures 3(a) and 3(b) show inverse compressibility as a function of density and magnetic field.  We observe distinct incompressible peaks corresponding to quantum Hall states at $\nu = \pm 4$ that extend all the way down to $B = 0$, where they merge into an incompressible region at the charge neutrality point [Figs. 3(a)-(c)].  Surprisingly, the $\nu = \pm 4$ gaps do not vanish at low fields; in fact, they increase with decreasing field below $0.2$ T, as shown in Fig. 3(d).  The zero-field incompressible peak has a characteristic width of approximately $10^{10}$ cm$^{-2}$ [Fig. 3(c)], and integration yields a step in chemical potential of nearly $2$ meV.  This increase in chemical potential is too large to be explained by a disorder-induced electric field, which would only lead \cite{castrev2007} to gaps on the order of $0.1$ meV for charge inomogeneity of $10^{10}$ cm$^{-2}$.  We observe qualitatively similar behavior at all positions along the flake \cite{supplement}.  It is worthwhile to note that our measurements give no indication of the negative incompressibility \cite{kus2008} at zero field, but are consistent with more recent predictions \cite{borghi2010}.
\begin{figure}
\includegraphics[width=85mm]{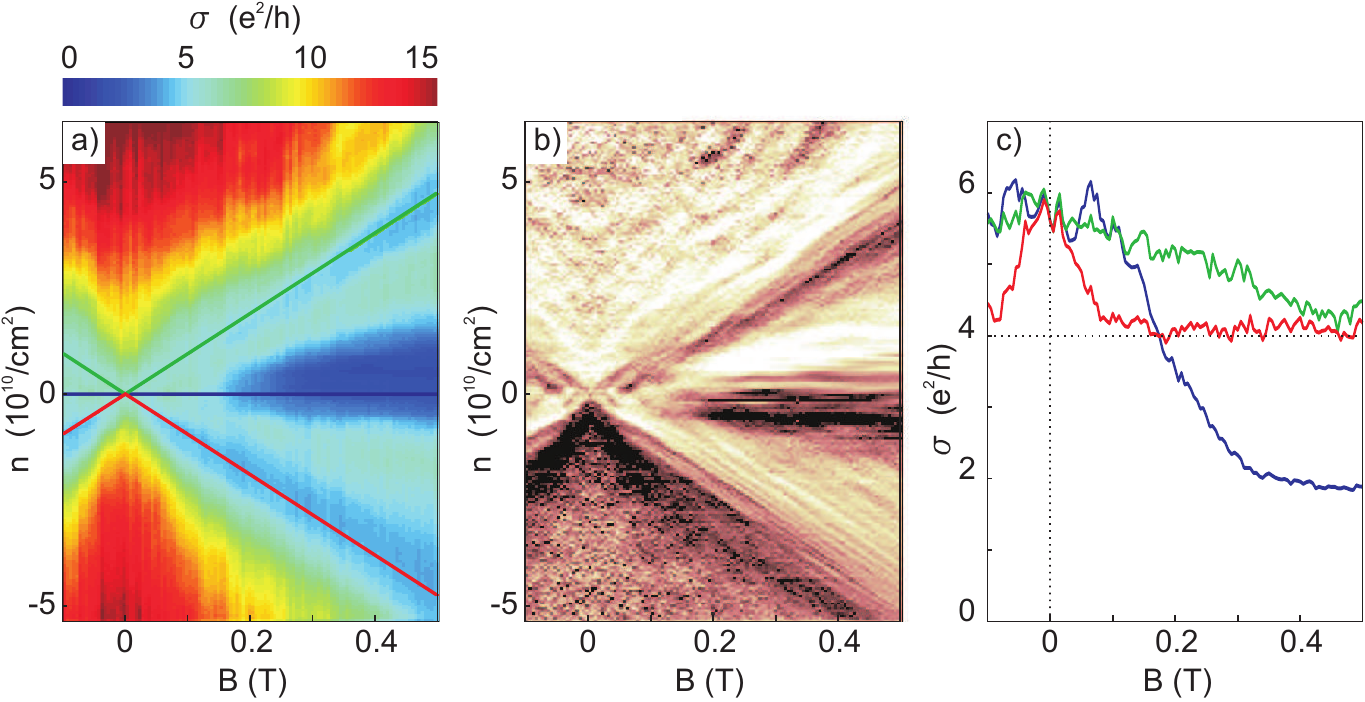}%
\caption{(Color online) (a) Two-terminal conductance as a function of carrier density and magnetic field. Quantum Hall plateaus at $\nu = \pm 4$ emerge at very low field, and the onset of the highly resistive state at $\nu = 0$ is also apparent. (b) Derivative of conductance with respect to carrier density.  Local peaks and minima with the same slope as $\nu = \pm 4$ are visible, indicative of localized states.  These localized states persist down to zero magnetic field.  (c) Conductivity along filling factors $\nu = - 4$ (red), $0$ (blue), and $4$ (green) as a function of magnetic field.}
\end{figure}

To further elucidate the origin of the low-field incompressible behavior, we have performed transport measurements on the same flake.  Two-terminal resistance as a function of carrier density and magnetic field is shown in Fig. 4(a).  Despite the relatively large jump in chemical potential that we observe in compressibility measurements, the resistance near the charge neutrality point at zero magnetic field is only a few k$\Omega$.  For comparison, we note that the gap size of the $\nu = 0$ state at $B = 0.5$ T is approximately $0.5$ meV, but even for this small gap, a marked decrease in conductance is already apparent [Fig. 4(c)].  Therefore, the formation of an energy gap with conduction mediated by activation is an insufficient explanation for the zero-field behavior that we observe.  The derivative of conductance with respect to carrier density is plotted in Fig. 4(b).  Several sharp lines with the same slope as the $\nu = 4$ state are apparent, and these conductance fluctuations are caused by localized states in the bilayer \cite{velas2010}.  The localized states, which indicate the presence of an energy gap, persist all the way to zero field.
\\
\indent
The incompressible behavior and transport characteristics at $B = 0$ indicate the presence of an interacting state.  In the limit of parabolic bands, the non-vanishing density of states at the charge neutrality point means that even infinitesimally small electron-electron interactions can lead to correlated states \cite{haldane88,raghu2008,min2008,nand2010,zhang2010,nandprl2010,vafek2010,lemonik2010,rahul2010}.  The existence of a zero-field incompressible peak and the fact that gap size along $\nu = \pm 4$ does not vanish at low fields are consistent with two proposed interacting states, which we discuss below.
\\
\indent
One proposal that is consistent with our measurements is the formation of an anomalous Hall insulator at low electric and magnetic field \cite{haldane88,raghu2008,min2008,nand2010,zhang2010,rahul2010}.  In such a state, time-reversal symmetry is spontaneously broken and domains form where the flake is at either $\nu = 4$ or $\nu = -4$.  It is reasonable that we observe incompressible behavior at both filling factors if we assume that the SET is too large to resolve individual domains.  In this scenario, the $B = 0$ conductance is dominated by edge state transport and should therefore remain at a few k$\Omega$ per square, consistent with our findings.  The fact that gap size along $\nu = \pm 4$ increases with decreasing field below about $0.2$ T [Figs. 3(a), 3(b) and 3(d)] can be understood to arise from the competition between the anomalous Hall phase and the standard quantum Hall gap as magnetic field is increased \cite{nand2010,rahul2010}.
\\
\indent
So far, we have assumed parabolic bands and neglected trigonal warping.  Trigonal warping modifies the band structure of bilayer graphene so that four Dirac cones emerge at low energies, leading to a $16-$fold degenerate LL at zero energy for low magnetic fields.  Effects of trigonal warping should be apparent \cite{mccannfalco2006} at densities as high as $10^{11}$ cm$^{-2}$, which is well within the experimentally accessible regime due to the low disorder in our sample.  Our results, however, cannot be explained by trigonal warping in the single-particle picture because the incompressible peaks at low magnetic field in Fig. 4(a) follow $\nu = \pm 4$ rather than $\nu = \pm 8$.  Recently, however, it was theoretically predicted \cite{vafek2010,lemonik2010} that electron-electron interactions can break rotational symmetry and modify the dispersion into a nematic phase that is characterized by two Dirac cones.  In this scenario, one would expect incompressible peaks along $\nu = \pm 4$, with energy gaps that scale as $\Delta_{\nu} \sim B^{1/2}$.  This means that gap size would remain relatively large at low fields, as we observe.  In this picture, the zero-field incompressible peak that we measure can be ascribed to the vanishing density of states, and because of the gapless nature of the spectrum, the conductance should remain relatively high, also consistent with our observations.

\begin{acknowledgments}
We would like to acknowledge useful discussions with L. S. Levitov, R. Nandkishore, S. Sachdev, M. S. Rudner and B. I. Halperin.  This work was supported by the U.S. Department of Energy, Office of Basic Energy Sciences, Division of Materials Sciences and Engineering under Award No. DE-SC0001819, by the 2009 U.S. Office of Naval Research Multi University Research Initiative (MURI) on Graphene Advanced Terahertz Engineering (GATE) at MIT, Harvard and Boston University, by Harvard's NSEC under National Science Foundation award No. PHY-0646094, and by the Alexander von Humbolt Foundation.  This work was performed in part at the Center for Nanoscale Systems (CNS), a member of the National Nanotechnology Infrastructure Network (NNIN), which is supported by the National Science Foundation under NSF award No. ECS-0335765. CNS is part of the Faculty of Arts and Sciences at Harvard University.
\end{acknowledgments}

\bibliography{PRLetal}

\end{document}



\title{EPAPS Document No. \#\#
\\
Supplementary Material and Methods for: Local Compressibility Measurements of Correlated States in Suspended Bilayer Graphene}


\author{Martin, J.}
\author{Feldman, B. E.}
\author{Weitz, R. T}
\author{Allen, M. T.}
\author{Yacoby, A.}
\affiliation{Department of Physics, Harvard University, Cambridge, MA 02138, USA}


\date{\today}

\pacs{73.22.Pr, 73.43.-f}
\keywords{graphene, quantum Hall effect}


\maketitle
\nopagebreak

\section{methods}
Sample fabrication was performed according to the method described in ref. \cite{feldman2009}.  Graphene was deposited onto a Si/SiO$_2$ wafer by mechanical exfoliation, and suitable bilayer flakes were contacted using an electron-beam lithography step followed by thermal evaporation of Cr/Au contacts.  Samples were then suspended using 5:1 buffered oxide etch, dried in a critical point dryer, and quickly transferred to the measurement system, an ultra-high vacuum scanning probe system with a $^3$He refrigerator.  All measurements were performed at base temperature of the system, $450$ mK.  Back gate voltage was limited to $\pm 15$ V to avoid collapse of the device.
\\
\indent
Scanning probe tips were fabricated using a fiber puller to generate a conical taper at the end of a quartz rod.  Al ($19$ nm) was thermally evaporated onto either side of the quartz rod to generate the two leads of the single electron transistor (SET).  Following a brief oxidation step, another $19$ nm of Al was evaporated onto the end of the conical tip to serve as the island of the SET.  For the measurements described in this paper, tip size was approximately $150 - 200$ nm and the tip was held between $50$ nm and $200$ nm above the flake.
\\
\indent
Compressibility measurements were performed using both AC and DC methods.  In the AC scheme, a small AC bias is applied to the back gate in order to weakly modulate the carrier density (by about $1.5 \cdot 10^9$ cm$^{-2}$) in the bilayer flake.  The corresponding changes in SET current are measured and converted to an electrostatic potential, with the conversion factor obtained by simultaneously monitoring the SET response to a $1$ mV AC potential applied directly to the sample.  In the DC scheme, the current through the SET is maintained at a fixed value using a feedback mechanism that modifies the potential of the sample.  The changes in sample potential provide a measure of the change in electrostatic potential caused by a given change in carrier density.  In both cases, $d\mu/dn$ and $\mu(n)$ are obtained because the electrostatic potential is related to the chemical potential of the flake by $e\Phi = \mu$.
\\
\indent
Due to the finite sample size, the SET is also affected in both schemes by fringing fields from the back gate \cite{martin2008,martin2009}, which give rise to a constant offset in $d\mu/dn$.  We account for this parasitic capacitance by subtracting a constant value so that $d\mu/dn = 0$ within each Landau level.  This is necessarily an overestimate, because it corresponds to a delta-function representation of the Landau level, whereas our Landau levels are broadened by disorder.  Therefore, our gap sizes should be taken as lower limits.  Incompressible peaks at each filling factor are fit with a single Lorentzian, with amplitude and width as fitting parameters.  The uncertainty in effective mass is calculated by adding in quadrature the uncertainty at each filling factor, as obtained by a linear regression.

\section{Zero-Field Compressibility at Large Densities}
\begin{figure}
\includegraphics[width=90mm]{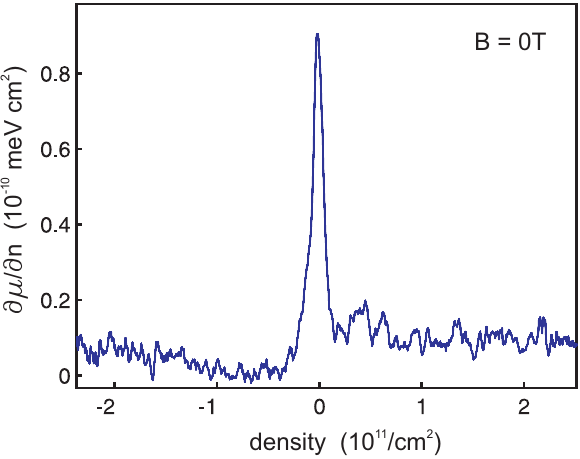}%
\caption{Inverse compressibility as
a function of carrier density at zero magnetic field. A sharp peak near the charge neutrality point
is apparent, consistent with the measurements presented in the main portion of the paper. There
is little variation in compressibility outside the central peak.\label{figS1}}
\end{figure}

Figure S1 shows the inverse compressibility of our sample at zero magnetic field, taken over a
larger density range and at a slightly different position along the flake. Similar to the
measurements presented in the main portion of the paper, a sharp incompressible peak is
apparent at the charge neutrality point. As electron density is increased beyond this central peak,
little change in compressibility occurs over the entire measurement range. For holes, inverse
compressibility increases slightly as density is increased. The origin of this behavior is unclear,
and the change is an order of magnitude smaller than the negative compressibility predicted by
Kusminskiy \textit{et. al} \cite{kus2008}. It also appears that the compressibility is somewhat larger for holes than for
electrons. Such behavior is consistent with the inclusion of the $\gamma_4$ hopping parameter, which
increases the density of states for holes and decreases it for electrons \cite{henriksen2010}. However, a non-negligible $\gamma_4$ would also be expected to cause an asymmetry between the energies of electron and
hole Landau levels, which we do not observe.

\section{Spatial Scans at Low Magnetic Field}
\begin{figure}
\includegraphics[width=160mm]{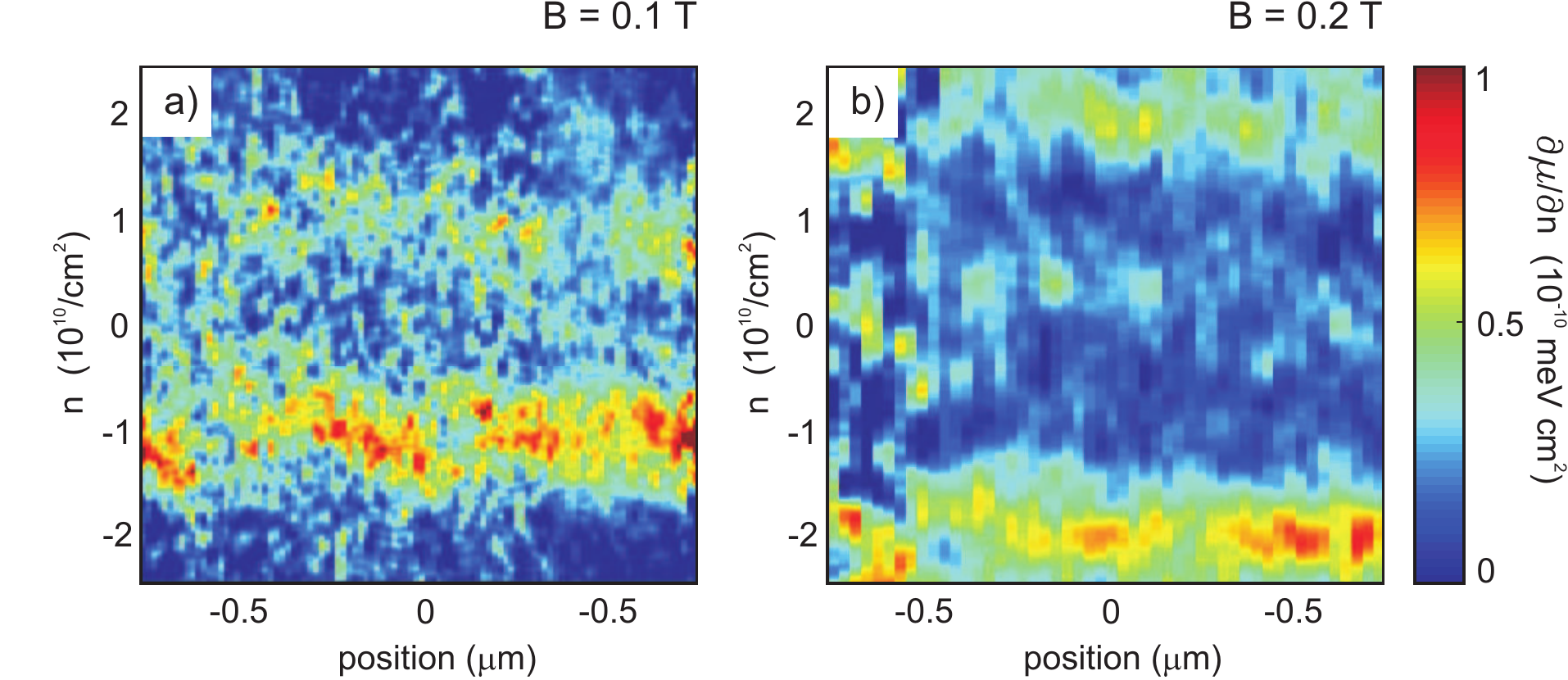}%
\caption{(a) Inverse compressibility as a
function of carrier density and position along the flake at $0.1$ T.
The incompressible peaks at $\nu = \pm 4$ (centered at approximately $\pm 10^{10}$ cm$^{-2}$) are consistently apparent, independent of position.
The incompressible peak on the hole side is consistently stronger than that on the electron side.
Some variations in the density at which the peaks occur is visible, providing an estimate of
disorder.
(b) Inverse compressibility as a function of carrier density and position along the flake at
$0.2$ T. The incompressible peaks at $\nu = \pm 4$ (centered at approximately $\pm 2 \cdot 10^{10}$ cm$^{-2}$) are again apparent along the entire flake. The electron-hole asymmetry is also apparent at this field, and
the $\nu = 0$ peak is just beginning to emerge.\label{figS2}}
\end{figure}
It is apparent in Fig. 3 that the incompressible peak height as well as the gap size are larger for
holes than for electrons along $|\nu | = 4$ at low magnetic fields. Figures S2(a) and S2(b) show
compressibility measurements taken as a function of density and position along the flake at $B =
0.1$ T and $0.2$ T, respectively. These measurements show that the electron-hole asymmetry is
consistent across the entire width of the sample. It is also worthwhile to note that the
fluctuations in density at which the incompressible peaks occur can provide another measure of the disorder in our samples. Consistent with estimates made by other means, it appears that the
charge inhomogeneity is less than $10^{10}$ cm$^{-2}$.

\section{Additional Measures of $\nu = \pm 4$ Gap Magnitude}
\begin{figure}
\includegraphics[width=140mm]{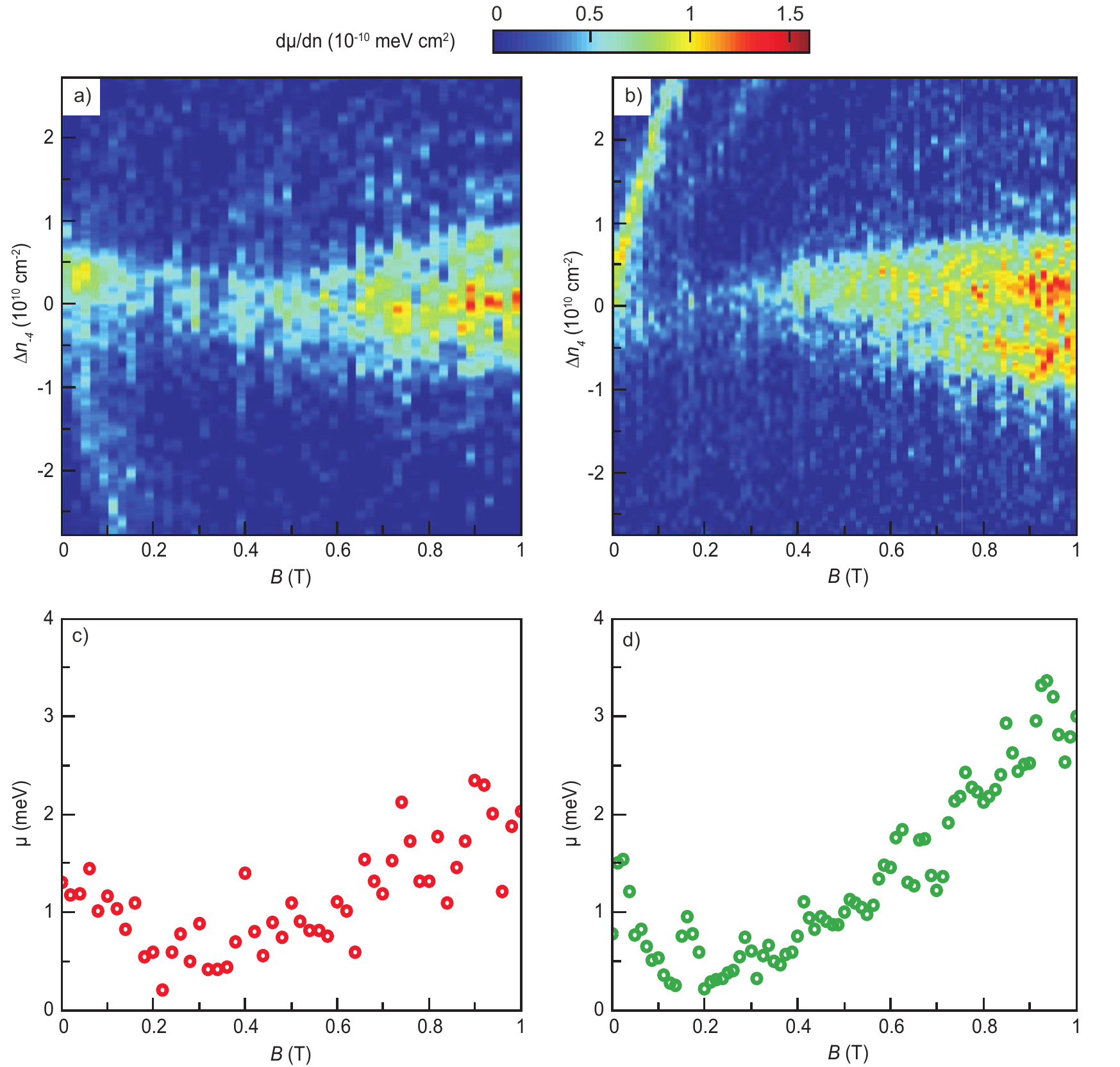}%
\caption{(a) Inverse compressibility
taken over a constant density range centered around $\nu = -4$ as a function magnetic field. Data
was taken at a slightly different position and height compared to that presented in the main
paper. At low fields, the $\nu = 4$ peak is visible as well.
(b) Inverse compressibility taken over a
constant density range centered around $\nu = 4$ as a function magnetic field. Data was taken at a
slightly different position and height compared to that presented in the main paper. At low
fields, the $\nu = 0$ and $-4$ peaks are visible as well.
(c) Gap size along $\nu = -4$ as a function of
magnetic field, as fit by a single Lorentzian. For $B < 0.1$ T, the gap size may be an overestimate
due the effect of the $\nu = 4$ incompressible peak.
(d) Gap size along $\nu = -4$ as a function of
magnetic field, as fit by a single Lorentzian. For $B < 0.1$ T, the gap size may be an overestimate
due the effect of the $\nu = 4$ incompressible peak.\label{figS3}}
\end{figure}
Figs. S3(a) and S3(b) show 'slanted' scans of inverse compressibility which follow a fixed carrier
density range surrounding filling factors $\nu = -4$ and $4$, respectively, at magnetic field between $0$
and $1$ T. At low fields, the second $|\nu | = 4$ peak and the $\nu  = 0$ peak are visible as well. These
measurements were performed at slightly different positions and heights compared to those
presented in the paper. Figs. S3(c) and S3(d) show the extracted gap size at each respective filling
factor as a function of magnetic field. We again observe that the gap size along $\nu = \pm 4$ does not
vanish at low field, but increases with decreasing field below $0.2$ T. For $B > 0.2$ T, gap size
increases linearly with field, as expected. The qualitative agreement between this dataset and
that presented in the main paper provides a second indication that our findings are not strongly
dependent on position.

\bibliography{PRLetal}